\documentclass[twocolumn,showpacs,nofootinbib,aps,prd]{revtex4}

\usepackage{graphicx,url,color}
\usepackage{amsmath,amsthm,amssymb}
\usepackage{multirow}
\usepackage{fmtcount} 
\renewcommand{\vec}[1]{\boldsymbol{#1}}

\newcommand{\beq}{\begin{equation}}
\newcommand{\eeq}{\end{equation}}

\newcommand{\D}{{\rm d}}
\newcommand{\unit}[1]{\ensuremath{\, \mathrm{#1}}}

\def\lsim{\raise0.3ex\hbox{$\;<$\kern-0.75em\raise-1.1ex\hbox{$\sim\;$}}}
\def\gsim{\raise0.3ex\hbox{$\;>$\kern-0.75em\raise-1.1ex\hbox{$\sim\;$}}}

\begin{document}

\title{Antideuteron Limits on Decaying Dark Matter with a Tuned Formation Model}

\author{L.~A.~Dal, A.~R.~Raklev$^1$}

\affiliation{$^1$Department of Physics, University of Oslo, Norway}

\begin{abstract}
We investigate the production of antideuterons from decaying dark matter, using gravitinos in supersymmetric models with trilinear R-parity violating (RPV) operators as an example. 
The model used for antideuteron formation is shown to induce large uncertainties in the predicted flux, comparable to uncertainties from cosmic-ray propagation models. 
We improve on the formation model by tuning hadronization and coalescence parameters in Monte Carlo simulations to better reproduce the hadron spectra relevant for antideuteron production. 
In light of current bounds on fluxes and future prospects from the AMS-02 and GAPS experiments we set limits on RPV couplings as a function of the gravitino mass. 
\end{abstract}


\maketitle

\section{Introduction}

The last few years have seen a lot of interest in decaying dark matter scenarios, in which dark matter (DM) candidates are very long lived particles.
Having decaying rather than annihilating DM candidates leads to important consequences for indirect detection signals, as the flux of decay products scale with the density of dark matter, 
whereas the flux of annihilation products scales with the DM density squared. 
Such long lived particles would be natural if the only coupling to Standard Model particles is very weak. 
Gravitational interaction is one such alternative.

In supersymmetric models, negative results from LHC searches for the standard missing energy signature of dark matter suggest looking at alternative models where R-parity is violated. 
With the gravitino as the lightest supersymmetric particle, the strength of the gravitational coupling means that gravitinos can naturally have lifetimes exceeding the age of the Universe. 
Such gravitino DM scenarios have been considered for some time now, beginning with~\cite{Takayama:2000uz,Buchmuller:2007ui,Lola:2007rw}, 
with specific indirect detection signatures of gamma-rays, positrons, antiprotons and neutrinos.

With a well motivated example of decaying dark matter, we will here investigate the production of antideuterons in gravitino decays for models with trilinear R-parity violating (RPV) operators. 
Use of the antideuteron channel for indirect detection of DM was first suggested in Ref.~\cite{Donato:1999gy}, and antideuterons have already been considered for generic decaying dark matter~\cite{Ibarra:2009tn}, 
and also specifically for gravitino DM~\cite{Grefe:2011dp} in models with bilinear RPV operators. 
We note several important differences between the present paper and previous works: 
The generic DM study~\cite{Ibarra:2009tn} includes only two-body decays with no baryon or lepton number violation, 
and therefore cannot describe the three-body decays of gravitinos into quarks and leptons due to  trilinear RPV operators. 
The work on RPV gravitino decays~\cite{Grefe:2011dp} does take three-body RPV decays into account, but only for models with bilinear RPV operators. We shall see that the baryon number violating trilinear couplings produce significantly more antideuterons than any other RPV decays previously considered. Moreover, we will use the resulting predicted antideuteron flux at Earth to set limits on the trilinear RPV couplings, and unlike~\cite{Ibarra:2009tn}, 
we do not make the simplifying assumption of isotropic and uncorrelated distributions of antiprotons and antineutrons in the decays. 

However, for readers interested in antideuteron limits for generic dark matter models, 
the most interesting novelty in this paper is probably the inclusion and treatment of uncertainties in the formation model for antideuterons, 
in particular hadronization effects. 
We investigate the effect of phenomenological hadronization parameters in the {\tt Herwig++} Monte Carlo event generator~\cite{Bahr:2008pv,Arnold:2012fq} and the $p_0$ parameter of the coalescence model, and re-tune these to better describe the particular hadron spectra relevant for antideuteron production. We present a set of recommended parameters for further studies that result in an improved description of antideuteron formation with smaller and quantifiable uncertainties. 

The trilinear RPV terms involved in the gravitino decays under study have the form
\beq
W\sim\lambda_{ijk} L_iL_j\bar{E}_k + \lambda'_{ijk}L_iQ_j\bar{D}_k+\lambda''_{ijk}\bar{U}_i\bar{D}_j\bar{D}_k,
\label{eq:WRPV}
\eeq
in the superpotential. Here $\lambda_{ijk}$, $\lambda'_{ijk}$ and $\lambda''_{ijk}$ are RPV couplings, $Q$ and $L$ are left-handed quark and lepton superfield doublets respectively, while $\bar{E}$, $\bar{U}$ and $\bar{D}$ are the corresponding left-handed superfield singlets for leptons and up and down type quarks. The first two terms violate lepton number, while the third violates baryon number conservation. From gauge symmetry, $i\neq j$ for $\lambda_{ijk}$ and $j\neq k$ for $ \lambda''_{ijk}$. For the present purpose of antideuteron production we are interested in the couplings that give antiquarks in gravitino decays: $\lambda'$ and $\lambda''$.

While these operators, if allowed, can cause problems, most seriously with proton decay, completely removing them by introducing R-parity~\cite{Fayet:1977yc} by hand seems excessive and {\it ad hoc}.
In particular it is known that the proton can be protected from decay by up to dimension--6 operators by introducing so-called trialities instead, disallowing either lepton or baryon number violation terms in Eq.~(\ref{eq:WRPV})~\cite{Ibanez:1991hv,Ibanez:1991pr,Lola:1993ip}.

Limits on trilinear couplings have been set on the basis of indirect detection experiments in the antiproton, positron and gamma-ray channels~\cite{Lola:2007rw,Bomark:2009zm}. 
These are generically much stronger than other (collider) limits on RPV couplings, resting only on the assumption that the gravitinos constitute a significant fraction of the total DM. 
The limits' dependence on sparticle masses is dominated by the gravitino mass; the other sparticles are less important, unless there is a dramatic difference in scales compared to the gravitino mass~\cite{Lola:2007rw}.\footnote{For the purposes of this paper we set all other sparticle masses to 1 TeV, just outside the current reach of the LHC.} 

As noted in~\cite{Ibarra:2012cc}, DM decay and annihilation involving final state quark--antiquark pairs are in general already severely constrained by antiproton searches. However, we will see that the trilinear operators we investigate here have stronger bounds from antideuterons due to the possibility of a triple antiquark final state. Antideuterons from DM decay or annihilation are also expected to have a much flatter spectrum at low energies than those from the main background, cosmic-ray--gas interactions. In conjunction with the extremely low level of background expected, this  leads to even stronger limits on the trilinear couplings. 

The present upper bound on the antideuteron flux that we use to set limits comes from the BESS balloon experiment~\cite{Fuke:2005it}, 
while the AMS-02 experiment running onboard the ISS \cite{AMSICRC30} and the planned GAPS balloon experiment~\cite{Koglin:2008zz} are expected to deliver stronger bounds in the near future. We give our predictions for future limits given null results from these searches.

The starting-point of the calculation of the expected cosmic-ray flux of antideuterons from gravitino decays is the decay width for gravitinos in the processes $\tilde G\to \nu_i d_j \bar d_k,\ell_i^- u_j \bar d_k$ and $\tilde G\to u_i d_j d_k$ found in \cite{Moreau:2001sr}. 
The parton level processes are showered and hadronized using the {\tt Herwig++ 2.6.0} Monte Carlo event generator~\cite{Bahr:2008pv,Arnold:2012fq}. 
The production of antideuterons from the resulting antiprotons and antineutrons is described by the so-called coalescence model. 
The complete formation model, from parton level to antideuterons, will be discussed in more detail in Sec.~\ref{sec:Coalescence}, including the uncertainties from the hadronization model used. 
The derived antideuteron spectrum is then propagated through the Galaxy, and the antideuteron flux at Earth is found in Sec.~\ref{sec:prop}. 
In Sec. \ref{sec:limits} we then use this flux to set limits on the relevant trilinear RPV couplings, and make predictions for what limits can be set with future experiments.

\section{Antideuteron production} \label{sec:Coalescence}

The fusion of antiprotons and antineutrons into antideuterons is commonly described by the coalescence model. 
In this model, any $\bar p\bar n$ pair with a momentum difference $\Delta p < p_0$, for some chosen value of $p_0$, will fuse to produce an antideuteron. 
In our calculation, this condition is evaluated in the center-of-momentum frame of the individual $\bar{p}$--$\bar{n}$ pairs. 
In the limit $m_{\bar{p}}=m_{\bar{n}}$, this is equivalent to the condition on invariant relativistic momentum used in {\it e.g.}\ Ref.~\cite{Ibarra:2012cc}.

Traditionally, the coalescence prescription has been applied to the averaged energy spectra of the antinucleons from a process, 
but this approach is based on the assumption that the antinucleon spectra are isotropic and uncorrelated.
These assumptions have been found not to hold in general in DM decay/annihilation processes~\cite{Dal:2012my,Kadastik:2009ts}, 
and the results presented here are therefore generated by applying the coalescence prescription on a per-event basis.

\subsection{Current coalescence models} \label{sec:p0}
The coalescence momentum $p_0$ can be found by running simulations of a particular process and fitting $p_0$ so that the result matches available experimental data. In principle, an uncertainty on the $p_0$ value can then be extracted and propagated to antideuteron fluxes.
A wide range of $p_0$-values have previously been used in the literature, however, the majority of these were found using the suboptimal isotropic assumption.

One of the major difficulties in reducing errors is the lack of 
data on antideuteron production at colliders to constrain models. In Refs.~\cite{Kadastik:2009ts,master,Dal:2012my} $p_0$ was calibrated against the single ALEPH antideuteron data point from hadron production at the Z-resonance~\cite{Schael:2006fd}, using per-event coalescence in {\tt Herwig++} and {\tt Pythia 8}~\cite{Sjostrand:2006za,Sjostrand:2007gs}.
The resulting values were $p_0 = 110$~MeV for {\tt Herwig++} and $p_0 = 162$~MeV for {\tt Pythia}, giving very different antideuteron spectra and indicating large model uncertainties of the order of a factor 2--4 in the observable flux~\cite{Dal:2012my}.

Recently, the authors of Ref.~\cite{Ibarra:2012cc} have pointed out that antinucleons produced in weak decays should not participate in antideuteron formation, as they would be produced too far away from the primary vertex to interact with antinucleons produced in hadronization or other decays.
This reduces the number of antinucleons available for antideuteron formation substantially,\footnote{In $e^+e^-$ collisions at the Z-resonance, we find the antinucleon production rate to be reduced by a factor $\sim1.5$, leading to a reduction in the number of antideuterons produced for a given value of $p_0$ by a factor $\sim2-3$.} and the values of the coalescence momenta required to match experimental data become correspondingly higher.
In~\cite{Ibarra:2012cc} $p_0$ is calculated  for several experimental datasets using per-event coalescence in {\tt Pythia}, and  best fit $p_0$ values are found in the range 133--236~MeV. In Table~\ref{tab:p0values} we compare these results to our own results with {\tt Herwig++} for the same experiments using the default hadronization parameters. 
\begin{table}[h!]
\begin{tabular}{|l|c|c|c|c|}
  \hline
	Experiment    & Process & {\tt Pythia 6} & {\tt Pythia 8} &{\tt Herwig++} \\ 
    \hline
    	ALEPH~\cite{Schael:2006fd}  				& $e^+e^-$  & --    & 192 & 159		\\
    	CLEO~\cite{Asner:2006pw}     				& $e^+e^-$ 	& -- 	& 133 & 145		\\
    	ZEUS~\cite{Chekanov:2007mv} 				& $ep$      & 236 	& --  & 150		\\
    	CERN ISR~\cite{Alper:1973my,Henning:1977mt} & $pp$      & --   	& 152 & 221		\\
    	ALICE~\cite{Sharma:2012zz}   				& $pp$      & 230  	& --  & 154		\\
    \hline
\end{tabular}
\caption{Best fit values of $p_0$ in MeV for various experiments and event generators. Values for {\tt Pythia} taken from~\cite{Ibarra:2012cc}. Note that not all processes are available in {\tt Pythia 8}.}
\label{tab:p0values}
\end{table}

The experiments where data are available is, as mentioned, $e^+e^-$  interactions at ALEPH, production from $\Upsilon(1S)$ decay at CLEO~\cite{Asner:2006pw}, production in $ep$--scattering at $\sqrt{s}=318$ GeV in ZEUS~\cite{Chekanov:2007mv}, $pp$-scattering at $\sqrt{s}=53$ GeV at the CERN ISR~\cite{Alper:1973my,Henning:1977mt} and $pp$-scattering at $\sqrt{s}=7$ TeV in ALICE~\cite{Sharma:2012zz}. We note a smaller spread in the fits using {\tt Herwig++}, with the exception of the CERN ISR data.

As discussed in Ref.~\cite{Dal:2012my}, the antideuteron production is quite sensitive to the hadronization procedure,  and it is therefore unlikely that the best fit $p_0$ values found by Ref.~\cite{Ibarra:2012cc} using versions of  {\tt Pythia} will be representative for events generated using {\tt Herwig++}. Differences between the Monte Carlo generators in Table~\ref{tab:p0values} indicate their different modeling of the underlying physics.

An interesting question that arises is whether the coalescence model should be sensitive to the process in which the coalescing antinucleons were produced. If not, one would {\it a priori} expect the different experiments to be more or less consistent with a common value of the coalescence momentum $p_0$. 
This is not immediately clear in Table~\ref{tab:p0values} for either event generator, although, as we shall see, the differences may be due to other parameters in the hadronization model. 
For an indication of whether the model is consistent across experiments we have performed a combined least squares fit of $p_0$ over all the experiments for {\tt Herwig++} using its default settings. This gives a best fit $p_0=152$~MeV with  $\chi^2 = 53.2$ for a total of 25 degrees of freedom (d.o.f.). The individual contributions to the $\chi^2$ are shown in Table~\ref{tab:chisqValues}, where we also show for comparison the $\chi^2$ for fits to the individual experiments.

\begin{table}[h!]
\begin{tabular}{|l|c|c|c|}
  \hline
	Experiment 	&  $\chi^2$, best fit $p_0$ & $\chi^2$, $p_0=152$~MeV & $N_{bins}$  \\ 
    \hline
    	ALEPH  	& 0.0  		& 0.2 	& 1		\\
    	CLEO    & 7.6 		& 10.5	& 5		\\
    	ZEUS 	& 3.7     	& 3.8	& 3		\\
    	CERN ISR& 5.0     	& 33.2	& 4+4	\\
    	ALICE   & 5.1     	& 5.5	& 9		\\
    \hline
\end{tabular}
\caption{$\chi^2$ contribution from the various experiments for the individual best fit values of $p_0$ and for the combined best fit value of $p_0 = 152$~MeV.}
\label{tab:chisqValues}
\end{table}

It is clear from Table~\ref{tab:chisqValues} that the individual experiments can be well described, but not when combined. 
One possible reason behind this failure is the choices made in the necessary phenomenological tuning of Monte Carlo event generators in order to reproduce data. Event generators such as {\tt Herwig++} and {\tt Pythia} are typically tuned with emphasis on particle multiplicities, and very little or not at all with respect to two-particle correlations. A systematic spread in the best fit value of $p_0$ between processes and energies is therefore to be expected because of the emphasis the various processes/experiments have been given in tuning, which in turn influences the predicted production of the antiproton and antineutron building blocks of the antideuteron.

The failure to describe data with statistical fluctuations means that the spread of $p_0$ values we have found are not a good measure of the uncertainty on the antideuteron flux coming from the coalescence model.
The authors of Ref.~\cite{Richardson:2012bn} have demonstrated how re-tuning hadronization parameters can be used to improve the physics description for a particular process (Higgs production), 
and to determine the underlying uncertainty from the tuning of these parameters. 
In the following section we will proceed by re-tuning a selection of the {\tt Herwig++} hadronization parameters in order to be able to simulate antideuteron production more consistently.

\subsection{Tuning of formation model}
\subsubsection{Choice of parameters}
The full set of parameters in the {\tt Herwig++} hadronization model constitutes a rather big parameter space, and a full parameter scan is not realistic with the large statistics required for per-event coalescence deuteron production. In addition to the coalescence momentum $p_0$, 
we therefore restrict ourselves to tuning the three parameters studied in Ref.~\cite{Dal:2012my} that were found to individually have the strongest impact on antideuteron production:\footnote{We do not include the parameter {\tt AlphaMZ}, 
the strong coupling at the $Z$-mass, which is constrained by high energy scattering data.}
{\tt PwtDIquark}, {\tt ClMaxLight} and {\tt PSplitLight}.

In order to make the role of these parameters clear, we briefly review the {\tt Herwig++} cluster hadronization model.
Following the perturbative QCD cascade, all gluons are first split into quark--antiquark pairs. Color-connected (di)quarks pairs are subsequently combined to form color singlet clusters.
Light clusters are decayed into hadrons, while clusters heavier than some mass threshold are iteratively fissioned into lighter clusters.
The {\tt ClMaxLight} parameter is one of two parameters involved in specifying this mass threshold for clusters consisting of light quarks ($u$, $d$ or $s$). The criterion for fission is that the cluster mass $M$ fulfills
\begin{equation}
M^p\ge m_C^p +(m_1+m_2)^p,
\end{equation}
where $m_1$ and $m_2$ the masses of the two partons in the cluster,  $m_C$ is {\tt ClMaxLight} and $p$ is the other parameter involved. Changing {\tt ClMaxLight} will affect the masses available in cluster decay and as a result the probability to create (anti)baryons.

For clusters above this threshold, a light quark--antiquark pair is popped from the vacuum, and two clusters are formed with one of the new and one of the original partons in each cluster. 
The inverse of {\tt PSplitLight} is a power that controls the mass distribution of the resulting clusters when they contain only light quarks.

The final step of the hadronization process is cluster decay. Given a cluster of flavour $(q_1,\bar{q}_2)$, a (di)quark--anti(di)quark pair $(q,\bar{q})$ or $(qq,\bar{q}\bar{q})$ is extracted from the vacuum, and two hadrons with the flavours $q_1\bar{q}$ and $q\bar{q}_2$, or $q_1qq$ and $\bar{q}\bar{q}\bar{q}_2$, are formed. The {\tt PwtDIquark} parameter controls the probability of choosing a diquark pair to be popped from the vacuum in this process, and thus directly controls the probability of creating baryons.

Hadronization parameters will in general affect both (anti)nucleon yields through the cluster masses and their two-particle correlations. For the parameters considered here, we have found that this holds true for {\tt ClMaxLight} and {\tt PSplitLight}, while {\tt PwtDIquark} appears to mainly affect the (anti)nucleon yields, which is to be expected since it controls the production of baryon--antibaryon pairs, as opposed to baryon or antibaryon pairs.

\subsubsection{Tuning procedure}
For the tuning of the formation model we use a least squares fit, minimizing the quantity
\beq
S = \sum_i^{\substack{\rm bins, \\ \rm experiments}} \left(\frac{y_i^{\rm exp} - y_i^{\rm MC}(\alpha_j)}{\sigma_i}\right)^2,
\label{eq:chisq}
\eeq
where $y_i^{\rm exp}$ and $y_i^{\rm MC}$ are the experimental and Monte Carlo values for bin $i$, respectively, $\alpha_j$ are the four parameters being tuned, and the sum is over all experimental data bins. 
In the fit we use all the available experimental data listed in Sec.~\ref{sec:p0} except the $pp$ data, and we also include data on proton production in $e^+e^-$ collisions (see below). 
Our reason for not including the $pp$-data is twofold: Firstly, generating sufficient statistics for these experiments is highly computationally expensive. 
Including CERN ISR and ALICE in our fit would increase the required CPU time by a factor of 6, and was thus not feasible in this context.
Secondly, the extremely large rise in the $\chi^2$ contribution from the CERN ISR data between the individual and the joint fit seen in Table~\ref{tab:chisqValues}
might indicate some process dependence in the coalescence model---even if this is not seen in the ALICE data---or that there might be some problem with the CERN ISR data or our interpretation of it.

The uncertainty $\sigma_i^2 = \sigma_{i,{\rm MC}}^2 + \sigma_{i,{\rm exp}}^2$ used for bin $i$ is given by  $\sigma_{i,{\rm MC}}$ and $\sigma_{i,{\rm exp}}$, the Monte Carlo and experimental uncertainties of the corresponding bin, respectively.
Ideally, one should generate enough events for the Monte Carlo uncertainty to be negligible compared to the experimental one, 
but for antideuterons, this is not practical in the context of a parameter scan due to the very low production rate. 
We therefore compromise by generating 100 times the number of events needed for the Monte Carlo uncertainty to match the experimental uncertainty when averaged over all bins.

In order to avoid tuning into unrealistic spectra for antiprotons and antineutrons--the antideuteron building blocks--we constrain the fit further by using antiprotons in the tune. We use antiproton data from the LEP experiments  OPAL~\cite{Akers:1994ez} and ALEPH~\cite{Buskulic:1994ft}. The OPAL data constitute half of the light baryon data used in tuning {\tt Herwig++}, the other half being proton production at similar energies in $e^+e^-$ collisions from SLD.
Because of the large number of data points available (26 antiproton data points from ALEPH alone compared to 9 antideuteron data points from ALEPH, CLEO and ZEUS combined) we re-weight the antiproton data in the fit with a factor 1/25 to keep it from dominating the parameter determination.

The antiproton data points of the two experiments are known to not be compatible at very high antiproton momenta, and should therefore not be used together directly. We combine the data of the two experiments using a weighted average
\beq
y_i = \left(\frac{y_{i,A}}{\sigma_{i,A}^2} + \frac{y_{i,O}}{\sigma_{i,O}^2} \right) \left( \frac{1}{\sigma_{i,A}^2} + \frac{1}{\sigma_{i,O}^2} \right)^{-1},
\eeq 
where the $A$ and $O$ subscripts denote ALEPH and OPAL, respectively.
We combine the experimental uncertainties accordingly, and use the differences in central values $\Delta y_i = |y_{\rm ALEPH} - y_{\rm OPAL}|$ as an additional systematic error describing our ignorance of the reason for the incompatible data, giving a combined uncertainty of
\beq
\sigma_i = \sqrt{\left( \frac{1}{\sigma_{i, A}^2} + \frac{1}{\sigma_{i, O}^2} \right)^{-1} + (\Delta y_i)^2}.
\eeq
As the energy bins of the two experiments do not match, we choose the binning of the ALEPH data set, and interpolate the OPAL data correspondingly.

For the minimization procedure we use the {\tt MIGRAD} algorithm in {\tt Minuit}~\cite{James:1975dr}. 
However, due to the per-event coalescence model used, it is prohibitively expensive in terms of computation time to directly evaluate the $\chi^2$ as many times as minimization 
requires (one parameter point takes $\sim 120$ CPU core hours on a modern processor).
Instead, we sample parameter points on grids, 
and use so-called radial basis functions (RBFs)---see {\it e.g.}\ Ref.~\cite{buhmann2003radial} and further discussion below---to approximate the shape of the function between the sampled points.
The global minimum is then found through an iterative procedure: first, we apply the {\tt MIGRAD} algorithm on the approximated function, starting in the best point sampled, to find a temporary global minimum.
We then evaluate this point with {\tt Herwig++} to find the true function value. 
If the difference between the true and the approximate function value at the minimum is greater than a chosen value, the RBFs are re-calculated with the newly evaluated point included, and the procedure is repeated.
We here accept the minimum if the difference between the true and the approximate function value is less than $\sim 0.5$, which is roughly the typical size of the errors on $S$ in our calculation due to limited statistics.

In the procedure to set up the RBFs, we sample an initial total of 342 points in the three {\tt  Herwig++} hadronization parameters in two grids with equal step sizes, shifted a half step with respect to each other in each parameter direction.
For each point in the other three parameters, we calculate the antideuteron flux for $p_0$-values in steps of 1~MeV from 5 to 400~MeV; this can be done per event and is computationally inexpensive. In the radial basis functions, however, we use only ten different values of $p_0$ per point in the other parameters. We do this to avoid bias from having a much larger number of samples in one parameter, and also because the calculation becomes highly memory intensive with a large number of points.

The step sizes and ranges fully covered by the two grids are listed in Table~\ref{tab:GridSetup}. We have also sampled regions outside the quoted ranges to investigate the possible existence of other global minima.
Note that not all regions of the parameter space are covered by both grids, but overlap between the grids is ensured in regions where the function value is low enough to allow for a possible global minimum.

\begin{table}[h!]
\begin{tabular}{|l|c|c|c|}
  \hline
	Parameter 			& Step size & Range, grid 1 & Range, grid 2	\\ 
    \hline
	$p_0$				& 10		& 110 -- 200	& 110 -- 200	\\
	{\tt ClMaxLight}  	& 1.33		& 2.00 -- 7.33	& 1.33 -- 5.33	\\ 
	{\tt PSplitLight} 	& 0.53		& 0.15 -- 2.82	& 0.42 -- 2.02	\\ 
	{\tt PwtDIquark}  	& 0.117		& 0.250 -- 0.717& 0.075 -- 0.775\\ 
    \hline
\end{tabular}
\caption{Step sizes and fully covered parameter ranges for the grid scans. $p_0$-values are in units of MeV.}
\label{tab:GridSetup}
\end{table}

\subsubsection{Radial basis functions}

A commonly used method for approximating function values between scattered multivariate data points is radial basis functions (RBFs), for a reference see Ref.~\cite{buhmann2003radial}.
Given a function $f(\vec{x})$ whose value is known at a set of distinct points $\vec{x}_i$, the goal is to construct an approximated function $s(\vec{x})$, such that
\beq \label{eq:RBFcrit}
 s(\vec{x}_i) =  f(\vec{x}_i).
\eeq
This can be done through a linear combination of radially symmetric functions $\phi(r)$ centered in $\vec{x}_i$,
\beq
  s(\vec{x})=\sum_i a_i \phi(\|\vec{x}-\vec{x}_i\|).
\eeq
The coefficients $a_i$ are found by imposing Eq.~\eqref{eq:RBFcrit}, and for the choices of $\phi(r)$ mentioned below, the solution is unique~\cite{buhmann2003radial}.
The functions $\phi(r)$ are referred to as radial basis functions, and popular choices include
\beq
\phi(r) = \left\{
	\begin{array}{ll}
		r  & \mbox{Linear} \\
		\exp(-c^2r^2) & \mbox{Gaussian} \\
		\sqrt{r^2+c^2} & \mbox{Multiquadratic}.
	\end{array}
\right.
\eeq
The two latter choices introduce a free parameter, $c$, that will affect the shape of $s(\vec{x})$.

For this work, we tested all of the RBF choices mentioned above, and found linear RBFs to be the best option.
We found that Gaussian and multiquadratic RBFs generally gave sharper minima than linear RBFs, but the widths of the minima depended significantly on the shape parameter $c$.
With no free parameters and somewhat wider minima, linear RBFs will thus give a less ambiguous and more conservative estimate of the parameter errors.

Since the RBFs are radially symmetric, the interpolation between points will depend on the normalization of the different parameters that we are trying to minimize over. 
The divergence of $\phi(r)$ is radially symmetric in the parameter space, but the data was sampled with different step sizes in the different parameter directions.
This means that the relative divergence is much larger in parameters with small step sizes than in parameters with large step sizes.
To avoid having any preferred directions in the interpolation, we therefore normalize the parameters such that the step sizes are equal in all parameter directions.

\subsubsection{Results}
\label{sec:fitresults}

In Table~\ref{tab:fitresults} we show the resulting values of the parameters at the minimum found for the $\chi^2$, $\chi^2_{\rm min}=10.6$ for effectively 14.2 d.o.f.\ due to the down-weighting of the proton data, and the $1\sigma$--error bands determined from the shape of the $\chi^2$ distribution.
\addtocounter{footnote}{1}
\footnotetext[\value{footnote}]{The $\chi^2$ around the minimum is highly non-parabolic, and the uncertainty was therefore calculated using the {\tt MINOS} algorithm in {\tt Minuit}~\cite{James:1975dr}. 
	This typically gives larger values for the uncertainties than calculating uncertainties from the error matrix.}

\begin{table}[h!]
\begin{tabular}{|l|c|c|c|}
  \hline
	Parameter 			& Default value	& Value at $\chi^2_{\rm min}$ 	& Uncertainty$^{\decimal{footnote}}$\\ 
	\hline
	$p_0$ 	       		&  -- 			&	143.2						& $ ^{+6.2}_{-5.5}$					\\[1ex]
	{\tt ClMaxLight}   	& 3.25 			& 	3.03						& $ ^{+0.18}_{-0.15}$				\\[1ex]
	{\tt PSplitLight}   & 1.20 			& 	1.31						& $ ^{+0.19}_{-0.32}$				\\[1ex]
	{\tt PwtDIquark}  	& 0.49 			& 	0.48						& $ ^{+0.15}_{-0.04}$  				\\[1ex]
	\hline
\end{tabular}
\caption{Results from hadronization parameter fit compared to default values in {\tt Herwig++}. $p_0$-values are in units of MeV.}
\label{tab:fitresults}
\end{table}

\subsection{Antideuteron spectra} \label{sec:Spectra}
Using the improved formation model and the uncertainties determined for its parameters in the previous section we now calculate the antideuteron spectra from gravitino decays.

We simulate a representative selection of trilinear RPV couplings and gravitino masses with potential final-state anti-quarks, both heavy and light anti-quarks, in the single dominant coupling approximation: $\lambda_{112}'$, $\lambda_{133}'$, $\lambda_{112}''$ and $\lambda_{323}''$. Our expectation is that the $\lambda_{112}''$ coupling with its three light (anti-)quarks will have the greatest (anti)deuteron production.

\begin{figure}[h!]
\includegraphics[width=0.5\textwidth]{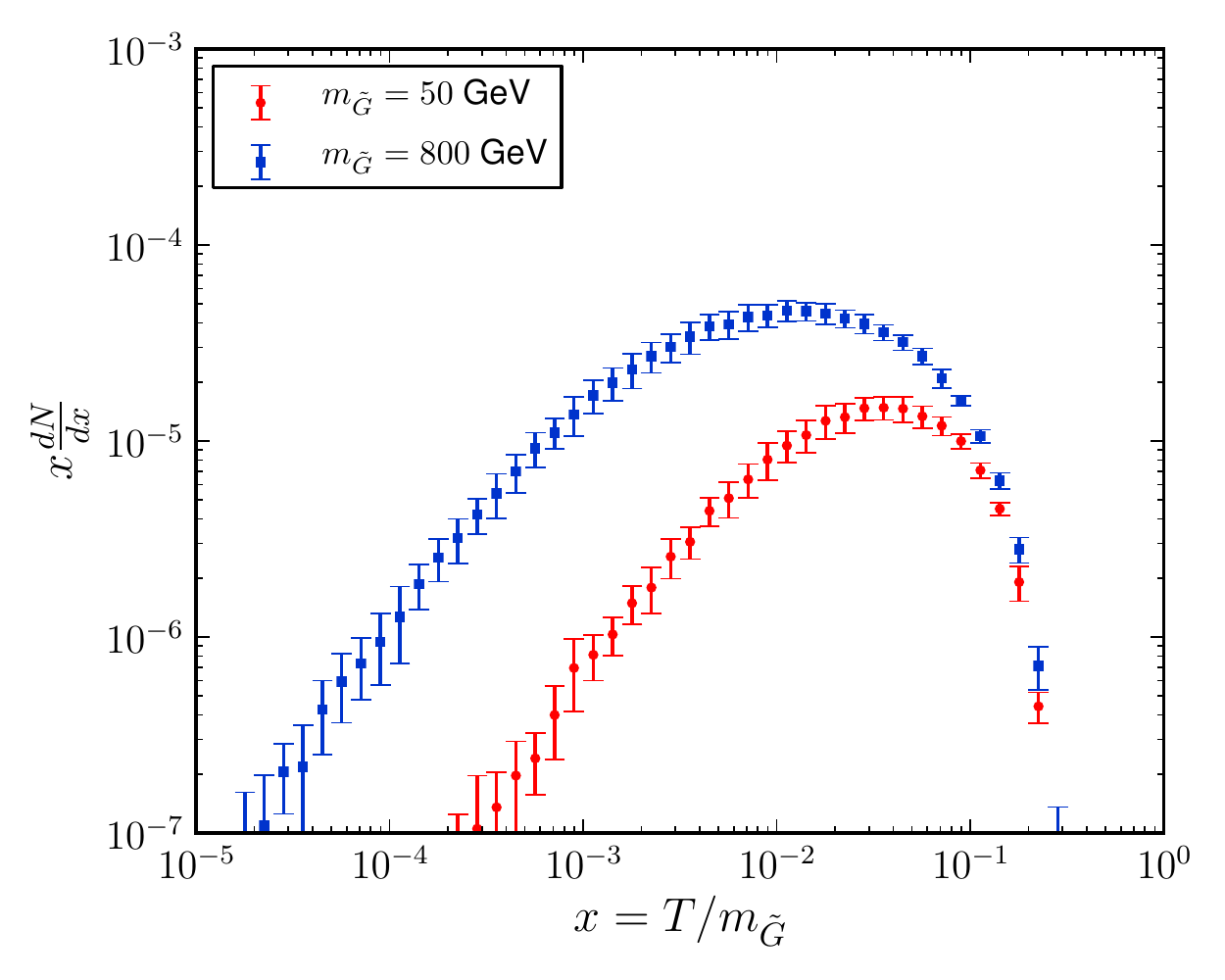}
\caption{Antideuteron spectra from gravitino decay through the $L_1Q_1\bar{D}_2$ operator for gravitino masses of 50~GeV and 800~GeV. For both masses $\lambda=1$ is used; the normalization of the spectra scales as $\lambda^2$.}
\label{fig:Gspec_LQD112}
\end{figure}

\begin{figure}[h!]
\includegraphics[width=0.5\textwidth]{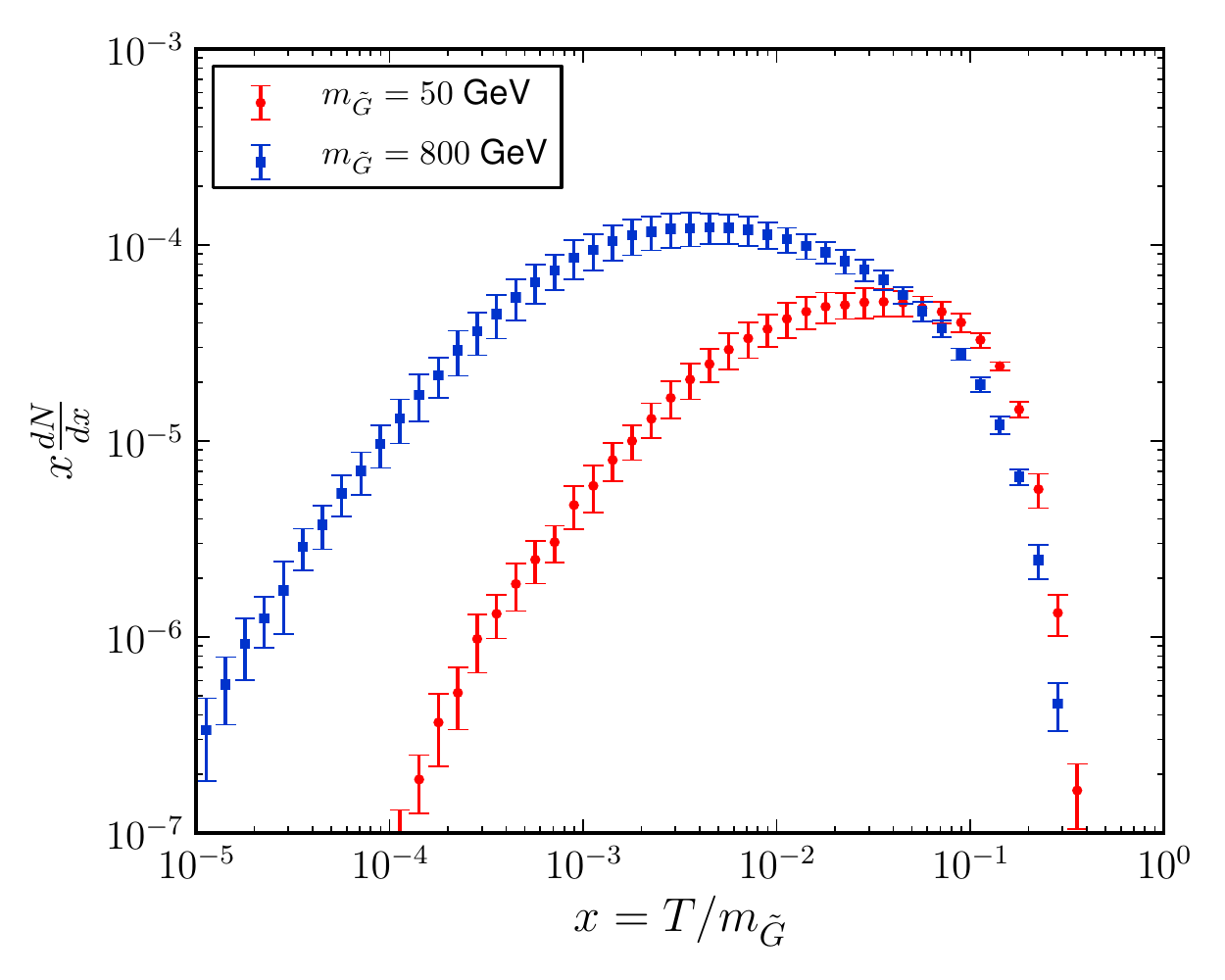}
\caption{Antideuteron spectra from gravitino decay through the $\bar{U}_1\bar{D}_1\bar{D}_2$ operator for gravitino masses of 50~GeV and 800~GeV. For both masses $\lambda=1$ is used; the normalization of the spectra scales as $\lambda^2$.}
\label{fig:Gspec_UDD112}
\end{figure}

In Figs.~\ref{fig:Gspec_LQD112} and~\ref{fig:Gspec_UDD112} we show the resulting antideuteron production spectrum for a selection of couplings and masses in terms of the scaled kinetic energy $x \equiv T/m_{\tilde{G}}$. 
The $1\sigma$ error-bars shown are determined by sampling ten sets of hadronization parameter values from a likelihood function constructed on the basis of the $\chi^2$--distribution found in the previous section. For each set of parameter values the antideuteron spectrum was calculated using $10^8$ events. As a result, the statistical uncertainties on the spectra are low compared to the uncertainty inferred from the parameter likelihood in all but the low energy ends of the spectra where the antideuteron yield is low.

As expected, decays through the $\bar{U}_1\bar{D}_1\bar{D}_2$ operator give a higher antideuteron yield than decays through $L_1Q_1\bar{D}_2$, but by less than an order of magnitude.
We find that the uncertainty on the spectrum contributed by $p_0$ and the investigated hadronization parameters to be a factor $\sim 1.3$ for most energies. This is substantially lower than the uncertainty from $p_0$ alone found in Ref.~\cite{Ibarra:2012cc} using various versions of {\tt Pythia} with default settings.

\section{Antideuteron flux at Earth}
\label{sec:prop}

\subsection{Astrophysical background}
One of the strengths of the antideuteron channel in DM searches is the very low expected astrophysical background flux.
Previous estimates of the background flux have been done by applying the coalescence model to antinucleon spectra under the assumption of isotropic antinucleon spectra.
More recently, however, a re-calculation of the astrophysical antideuteron flux was performed in Ref.~\cite{Ibarra:2013qt} using the more correct per-event coalescence in Monte Carlo event generators.
This leads to an expected antideuteron background that is a factor $\sim 2$ lower than the previous estimates.
The antideuteron background spectra used in this work are the re-calculated ones of Ref.~\cite{Ibarra:2013qt}.

\subsection{Gravitino decay signal}
\subsubsection{Propagation of antideuterons}
In order to find the resulting antideuteron flux at the Earth, the antideuterons must be propagated through the Galaxy from their point of origin in the DM halo.
Charged particles propagating through the Galaxy scatter on fluctuations in the turbulent magnetic field, leading to a random walk behaviour.
This movement is well described using a diffusion approximation, and here we use the so-called two-zone propagation model. This is a cylindrical diffusion model consisting of a magnetic halo of radius $R=20$~kpc and half-height $L$ (a free parameter), 
and a thin gaseous disk of the same radius and a half-height of $h = 100$~pc.

Assuming steady state conditions, and neglecting reacceleration and non-annihilating inelastic scattering, the diffusion equation describing this model is given by
\beq \label{eq:Diffusion}
 -D(T) \nabla^2 f  +  \frac{\partial}{\partial z} ( {\rm sign}(z) f V_c) = 
 Q - 2 h \delta (z) \Gamma_{\rm ann}(T) f \,,
\eeq
where $f(\vec{x},T)=dN_{\bar{d}}/dT$ is the number density of antideuterons 
per unit kinetic energy, $D(T)= D_0 \beta \mathcal{R}^\delta$  the 
(spatial) diffusion coefficient, $V_c$ a convective wind perpendicular to the Galactic disk, $z$ the vertical coordinate, $\beta=v/c$ the velocity and
$\mathcal{R}$ the rigidity of antideuterons in GV.  Here $\delta$, $D_0$, and $V_c$ are free parameters.

\begin{figure}
\includegraphics[width=0.5\textwidth]{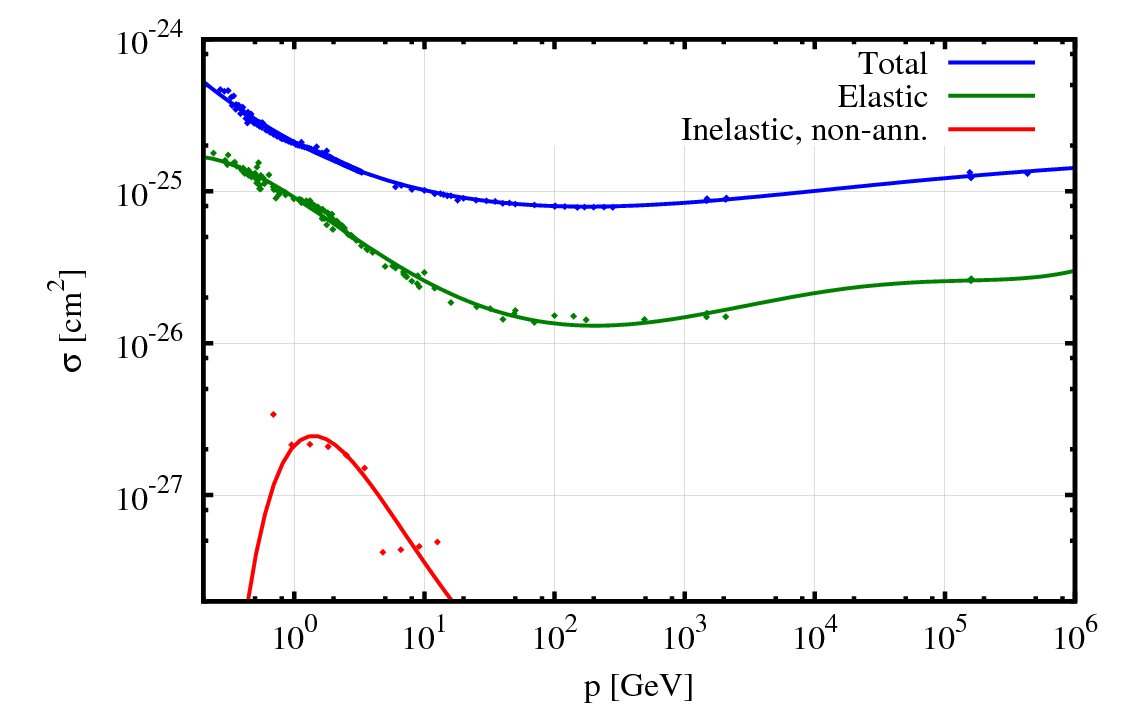}
\caption{Cross section data for antideuterons on interstellar protons as a function of the antideuteron momentum. The points indicate experimental data, while the lines show the fits to the data that were used in our calculations.}
\label{fig:ISM_XS}
\end{figure}

The annihilation rate, $\Gamma_{\rm ann} $, of antideuterons on interstellar gas in the Galactic disk is given by
\beq
\Gamma_{\rm ann}  = (n_{H} + 4^{\frac{2}{3}} n_{\rm {He}}) 
                  \langle \sigma^{\rm {ann}}_{\bar{d}p}v\rangle\, ,
\eeq
where we use the values $n_H \approx 1 \unit{cm^{-3}}$ and $n_{He} \approx 0.07 n_H$ for the number densities of H and He in the disc, respectively, 
and where the factor $4^{\frac{2}{3}}$ accounts for the difference in annihilation cross section between H and He, assuming simple geometrical scaling.
For the annihilation cross section, we use a fit to experimental data~\cite{Amsler:2008zzb,Schopper1988}, as seen in Fig.~\ref{fig:ISM_XS}.

The source term, $Q$, is determined by the dark matter density profile of our galaxy.
Most of the commonly used profiles can be parametrized as
\begin{equation}
 \rho(r) = \frac{\rho_0}{(r/a)^{\gamma}\left[ 1 + \left(r/a \right)^{\alpha} \right]^{(\beta - \gamma)/\alpha}} \,,
\end{equation}
where $a$ and $\rho_0$ are a characteristic length and a characteristic density, respectively, while $\alpha$, $\beta$ and $\gamma$ are dimensionless parameters.
The Navarro-Frenk-White (NFW) profile~\cite{Navarro:1995iw} is given by $\alpha=1$, $\beta=3$, $\gamma=1$.

For the free parameters in the model we adopt the three sets of values found in Ref.~\cite{Donato:2003xg} to yield maximal, 
median and minimal antiproton fluxes from DM annihilations, while being compatible with the observed B/C ratio.
These parameter sets are labeled 'max', 'med' and 'min' respectively, and their values are listed in Table~\ref{tab:PropModels}.

\begin{table}
\begin{tabular}{|l|c|c|c|c|}
  \hline
	Model 		& $L$ in kpc	& $\delta$	& $D_0$ in kpc$^2$\,Myr$^{-1}$	& $V_c$ in km\,s$^{-1}$	\\ 
    \hline
    	max 	& 15 			& 0.46 		& 0.0765						& 5							\\
    	med		& 4				& 0.7		& 0.0112						& 12						\\
		min		& 1				& 0.85		& 0.0016						& 13.5						\\
    \hline
\end{tabular}
\caption{Propagation parameters for the max, med and min models.}
\label{tab:PropModels}
\end{table}

The diffusion equation~\eqref{eq:Diffusion} can be solved (semi)analytically~\cite{Donato:2001ms}, giving an antideuteron flux at the position of the Earth for decaying DM~\cite{Ibarra:2009tn}
\beq \label{eq:FluxEarth}
 \Phi_{\bar{d}}(T,\vec{r}_{\odot}) = 
 \frac{v_{\bar{d}}}{4 \pi} \frac{1}{\tau} \frac{\rho_{\odot}}{M_{\rm DM}} R(T)
 \frac{dN_{\bar{d}}}{dT}  \,,
\eeq
where
\beq \label{eq:RT}
R(T)=\sum^\infty_{n=1} J_0 \left(\zeta_n \frac{r_{\odot}}{R} \right) \exp\left(-\frac{V_c L}{2K}\right) \frac{y_n(L)}{A_n \sinh(S_nL/2)},
\eeq
\beq \begin{split} \label{eq:yT}
y_n(Z) = &\frac{4}{J_1^2(\zeta_n)R^2}\int^R_0 \D r\ r J_0\left(\frac{\zeta_n r}{R}\right) \int^Z_0 \D z\ \Big\{ \\ 
	 & \exp\left(\frac{V_c(Z-z)}{2D}\right)\sinh\left(\frac{S_n(Z-z)}{2}\right)\frac{\rho(r,z)}{\rho_\odot} \Big\},
\end{split} \eeq
\beq \label{eq:An}
A_n=2h\Gamma_{\rm ann} +V_c +DS_n \coth(S_n L/2),
\eeq
and
\beq \label{eq:Sn}
S_n=\sqrt{\frac{V_c^2}{D^2}+4\frac{\zeta^2_n}{R^2}}.
\eeq 
The function $R(T)$ encodes the astrophysics of the propagation, and is completely independent of the particle physics of the DM decay. 
The function can thus be pre-calculated for any given set of propagation parameters and halo model, and subsequently used for any DM candidate decaying into antideuterons.
Solar modulations are further taken into account by replacing the final kinetic energy of the particles $T$ with a modified kinetic energy near the Earth~\cite{Gleeson:1968zza}, 
$T_\otimes=T-|Ze|\phi_{\rm Fisk}$, where the so-called Fisk potential $\phi_{\rm Fisk}=0.5$\,GV is an effective potential that parametrizes the energy loss from the solar wind.
The corresponding antideuteron flux near Earth is then given by
\beq
\Phi_\otimes = \frac{p^2_\otimes}{p^2} \Phi = \frac{2m_{\bar{d}}T_\otimes + T^2_\otimes}{2m_{\bar{d}}T + T^2} \Phi .
\eeq

\begin{figure}
\includegraphics[width=0.5\textwidth]{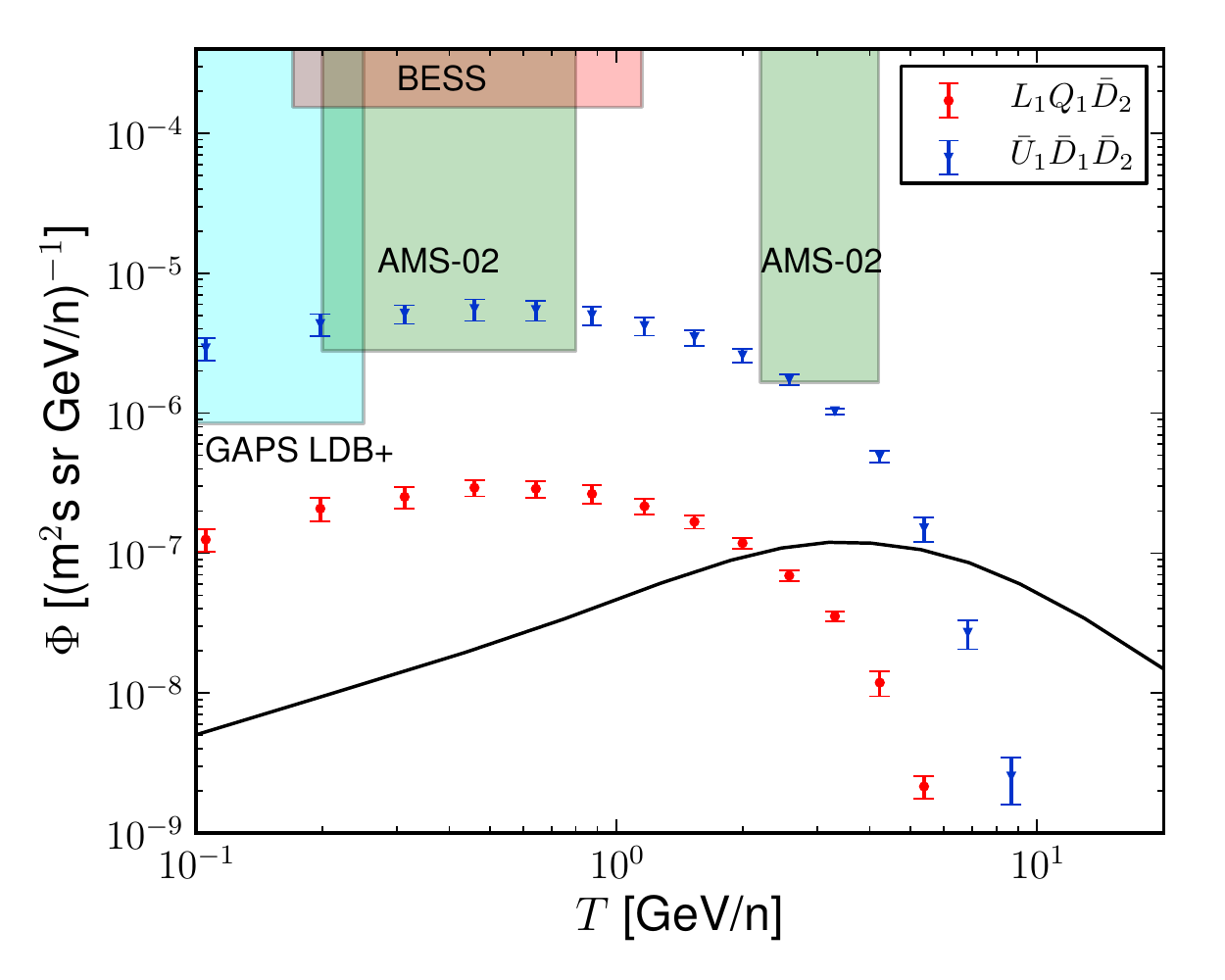}
\caption{Antideuteron flux at the Earth from gravitino decays with $m_{\tilde{G}}=50$~GeV for the NFW DM halo profile and 'med' propagation parameters. The colours show different RPV couplings. For all couplings $\lambda=10^{-5}$ is used; the normalization of the spectra scales as $\lambda^2$. The solid black line shows the expected background flux.}
\label{fig:Gflux50}
\end{figure}

\begin{figure}
\includegraphics[width=0.5\textwidth]{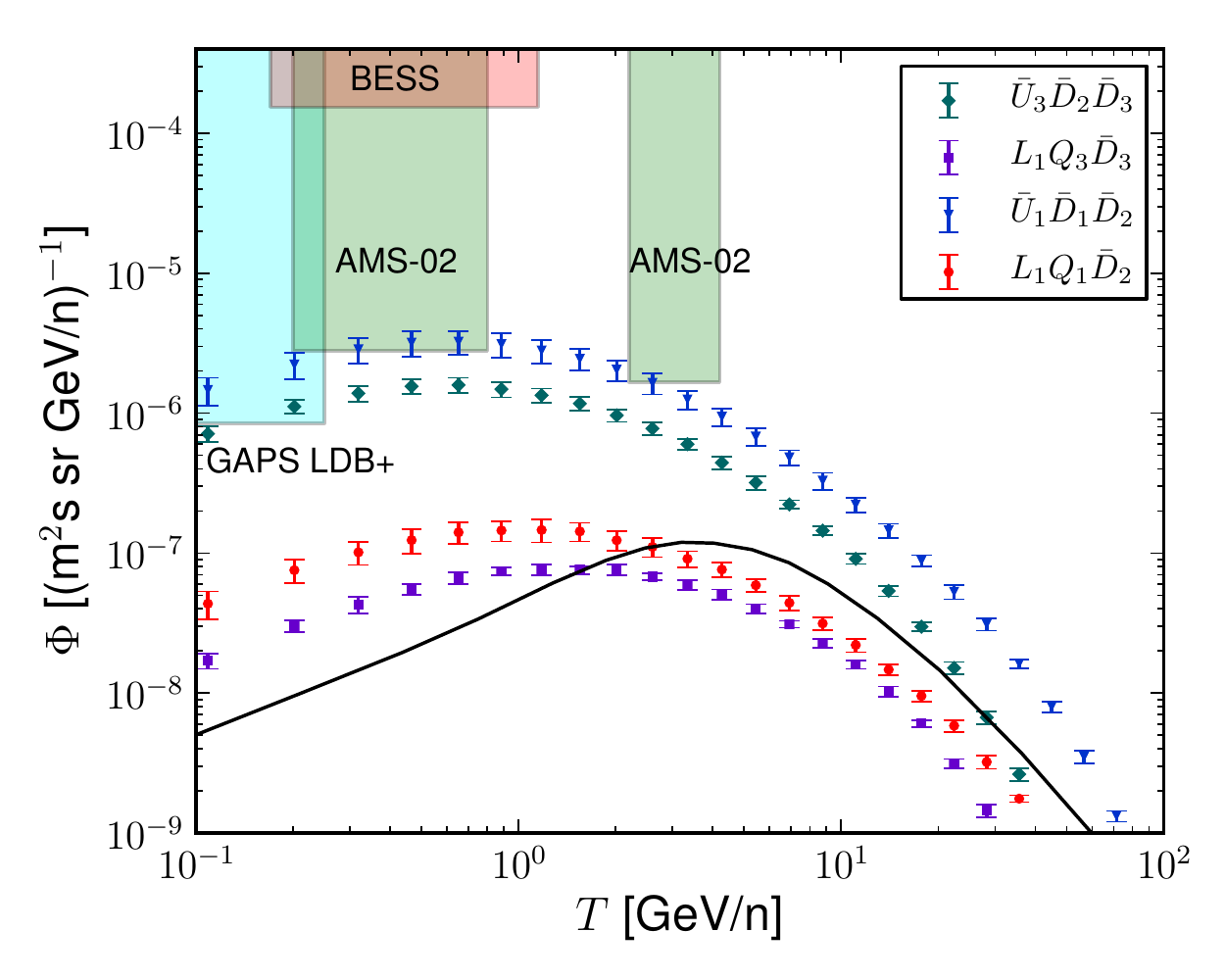}
\caption{Antideuteron flux at the Earth from gravitino decays with $m_{\tilde{G}}=800$~GeV for the NFW DM halo profile and 'med' propagation parameters. The colours show different RPV couplings. For all couplings $\lambda=10^{-9}$ is used; the normalization of the spectra scales as $\lambda^2$. The solid black line shows the expected background flux.}
\label{fig:Gflux800}
\end{figure}

With this diffusion model, solar modulation, and the antideuteron spectra calculated above, we show the resulting flux at the Earth for our choice of models in Figs.~\ref{fig:Gflux50} and~\ref{fig:Gflux800}. The error bands shown are determined  as in Sec.~\ref{sec:Spectra}.
RPV operators with similar flavour contents give similar fluxes, and are thus omitted here. 
An example is the $L_1 Q_2 \bar{D}_2$ operator, which yields a similar flux as the $L_1 Q_1 \bar{D}_2$ operator.

The BESS, AMS-02 and GAPS limits shown are (prospective) 95\% CL exclusion limits\footnote{The limits were calculated for the case of zero observed events in the GAPS and BESS experiments and in the TOF (low energy) region of the AMS-02 experiment. In the higher energy RICH region of the AMS-02 experiment, one observed event was assumed.} 
calculated using the $CL_s$ method, as discussed in Sec.~\ref{sec:limits}. 
We note that the fluxes near Earth from $LQ\bar{D}$ and $\bar{U}\bar{D}\bar{D}$ operators differ by an additional half order of magnitude compared to the corresponding source spectra seen in 
Figs.~\ref{fig:Gspec_LQD112} and~\ref{fig:Gspec_UDD112}.
This additional half order of magnitude comes from a corresponding difference in the gravitino lifetime. 

\subsection{Antideuteron detection experiments}
In order to set exclusion limits on the antideuteron flux and the RPV couplings, we need to be able to calculate the number of expected events at the various experiments that search for antideuterons. These calculations  will be described in the present section.

\subsubsection{BESS}
In the calculation of their limit on the cosmic ray antideuteron flux, the BESS collaboration uses the estimate~\cite{Fuke:2005it}
\beq \label{eq:fluxBESS}
\Phi_{\rm max} = \frac{N_{\rm obs}}{|S\Omega\ {\mathcal E}_{\rm tot}(1-\delta_{\rm sys})|_{\rm min} t_{\rm live} (T_2 - T_1)},
\eeq 
where $N_{\rm obs}$ is the number of observed events, $S\Omega$ is the geometrical acceptance, ${\mathcal E}_{\rm tot}$ is the total detection efficiency, $\delta_{\rm sys}$ is the total systematic uncertainty, 
$t_{\rm live}$ is the live time of the experiment, and $(T_1,T_2)$ is the kinetic energy range in which the flux is being measured. 
The 'min' subscript indicates that the minimal value of the quantity, as a function of energy, should be used to obtain a conservative estimate for the flux. This is also the reason for including the systematic uncertainty in the equation.
The corresponding equation for calculating the expected number of events for a given flux is
\beq \label{eq:nBESS}
N = \int_{T_1}^{T_2} S\Omega\ {\mathcal E}_{\rm tot}(1-\delta_{\rm sys}) t_{\rm live}\ \Phi\ dT .
\eeq 

For BESS the relevant experimental energy range is: $T_1 = 0.17$~GeV/n and $T_2 = 1.15$~GeV/n and the total systematic uncertainty was estimated by the experiment to lie at the 10\% level, $\delta_{\rm sys} = 0.1$. We take the  effective exposure $S\Omega\ {\mathcal E}_{\rm tot} t_{\rm live}$, as a function of energy, from Fig.~2 in~\cite{Fuke:2005it}.

\subsubsection{AMS-02} \label{sec:AMS}
The antideuteron analysis of AMS-02 has been described in detail in~\cite{Giovacchini:2007}, however, since this was published, the decision was made to replace the superconducting magnet of the experiment with a permanent magnet. The AMS collaboration has yet to publish a re-analysis using the new setup, and thus the best estimate possible will be to broadly follow the existing analysis.

We calculate the expected number of observed antideuteron events at AMS-02 through
\beq \label{eq:nAMS}
N_{\bar{d}} = \int A_{\bar{d}} C_{\rm Geo}t \Phi_{\bar{d}}\ dT ,
\eeq
where $t$ is the exposure time, $\Phi_{\bar{d}}$ is the antideuteron flux outside the Earth's magnetosphere, $ A_{\bar{d}}$ is the antideuteron acceptance with selection cuts taken into account, 
and $C_{\rm Geo}$ is the geomagnetic transmission function (GTF), which accounts for the drop in low energy charged particle fluxes due to the geomagnetic field.
The integral is performed over the sensitivity range(s) of the experiment. 
AMS-02 has two relevant sensitivity ranges, 0.2--0.8~GeV/n and 2.2--4.2~GeV/n, corresponding to the Time-of-Flight (TOF) detector and the Ring Imaging Cherenkov Counter (RICH), respectively.
We calculate the expected number of events separately for the two energy ranges.

\begin{figure}[h!]
\includegraphics[width=0.5\textwidth]{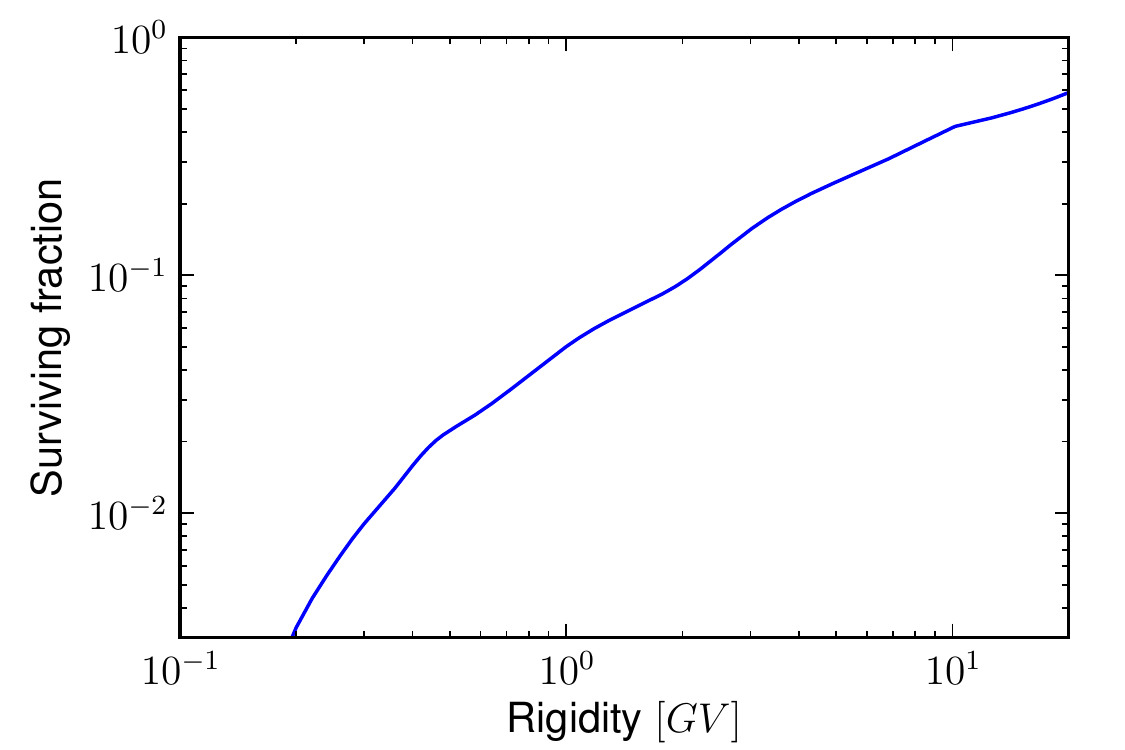}
\caption{Geomagnetic transmission function for the International Space Station orbit in quiet geomagnetic weather conditions. The GTF gives the surviving fraction of incident charged particles as a function of their rigidity.}
\label{fig:GTF}
\end{figure}

As in the original analysis, we assume an exposure time of 3 years. The acceptance, $A_{\bar{d}}$, after selection cuts, as a function of the kinetic energy, is taken from Fig.~2 in Ref.~\cite{AMSICRC30}.
For the geomagnetic transmission function, $C_{\rm Geo}$, we use the CREME96~\cite{CREME96_1997} precalculated GTF for the International Space Station orbit in quiet geomagnetic weather conditions, as shown in Fig.~\ref{fig:GTF}, as a function of particle rigidity. 

We note that this GTF yields a more conservative result than the GTF used in the original analysis. 
For comparison, the antideuteron flux from DM model b) in~\cite{Giovacchini:2007} was reported to yield two expected events in the TOF region with 3 years of exposure, while in our calculation this model only yields 1.2 events.

In the BESS experiments, no events were observed, and the expected number of background events is not needed for setting exclusion limits with the $CL_s$ method. If an event is observed, however, the expected number of background events is needed. The number of astrophysical antideuteron background events for AMS-02 can be calculated directly from Eq.~\eqref{eq:nAMS}.  However, the main source of background for AMS-02 is mis-reconstructed (anti)protons, deuterons and electrons. We estimate the number of such background events using
\beq \label{eq:nBG_AMS}
N_{\rm BG} = \sum_i \int  \frac{A_i}{R_i} C_{\rm Geo}t\Phi_{i}\ dT ,
\eeq
where $\Phi_i$ and $A_i$ are the flux and acceptance for particle species $i$, and $R_i$ is the corresponding rejection factor.

The rejection factor for a given particle species is defined as the ratio between the number of particles surviving the selection cuts and the number of particles (wrongfully) identified as antideuterons.
The rejection factors are in principle energy dependent, but only averaged rejection factors are publicly available. The full list of rejection factors, taken from~\cite{Giovacchini:2007}, is found in Table~\ref{tab:RejectionFac}.

\begin{table}
\begin{tabular}{|l|c|c l|}
  \hline
	Rejection Factor 			& TOF Range 				& \multicolumn{2}{c|}{RICH Range}						\\ 
    \hline
    $R_{\bar{p}}$  &  $6 \times 10^{6}$ 		& 	$2 \times 10^{5}$ 	& 	$T < 7.2~{\rm GeV}$ 		\\
								&							&	$2 \times 10^{4}$	&	$ 7.2 < T < 8.9~{\rm GeV}$ 	\\
    $R_p$  	        &  $8 \times 10^{11}$    	&	$8 \times 10^{11}$	&								\\
    $R_d$  	        &  $1.5 \times 10^{9}$    	&	$1.5 \times 10^{9}$	&								\\
    $R_e$  	        &  $2 \times 10^{9}$    	&	$5 \times 10^{7}$	&								\\
    \hline
\end{tabular}
\caption{AMS-02 rejection factors for antiprotons, protons, deuterons and electrons in the TOF and RICH energy ranges.}
\label{tab:RejectionFac}
\end{table}

The acceptances for the background species after antideuteron selection cuts, $A_i$, have not been released (except for deuterons), and in our analysis we therefore use the signal (antideuteron) acceptance $A_{\bar{d}}$ as a conservative estimate. For the fluxes of the background species, we use the fits to experimental data compiled in Fig.~2.4 of Ref.~\cite{vonDoetinchem:2009wc}.

Our calculation yields an expected background of 0.05 events in the TOF sensitivity range and 0.63 events in the RICH sensitivity range in the 3 year exposure. When we later calculate prospective limits from AMS-02, we assume zero observed events in the TOF range and one observed event in the RICH range; this in order to represent the most likely background-only observation.

\subsubsection{GAPS}
There is currently not enough information available in order to make an analysis for GAPS that is as thorough as the AMS-02 analysis. In order to estimate the number of expected events at GAPS, we will instead follow the approach of Ref.~\cite{Fornengo:2013osa}, where an effective exposure $\langle{\mathcal E}\rangle$ is estimated from
\beq \label{eq:nGAPS}
D = \int \Phi(E) {\mathcal E}(E) dE \simeq \Phi_{\Delta E} \times \langle{\mathcal E}\rangle \times \Delta E,
\eeq
where $D$ is the expected number of signal events, ${\mathcal E}(E)$ is the true, energy dependent detector exposure, and $\Phi_{\Delta E}$ is the expected  flux in the energy interval $\Delta E$.
For the GAPS Long Duration Balloon flight (LDB+), $D=1$ expected events correspond to a flux of $\Phi_{\Delta E} = 2.8\times10^{-7}$~(m$^2$ s sr GeV/n)$^{-1}$ in the energy interval 0.1--0.25~GeV/n~\cite{Fornengo:2013osa},
thus giving an effective exposure of $\langle{\mathcal E}\rangle = 2.36\times 10^7$~m$^2$~s~sr.

Using this estimated effective exposure, we can use Eq.~\eqref{eq:nGAPS} to find  the expected number of signal events for any other antideuteron flux. Unfortunately, we do not have an estimate for the number of background events, and we can therefore only set  limits for the case of zero observed events.

\section{Limits on RPV couplings}
\label{sec:limits}
With the expected signal and background fluxes in hand we can now proceed to setting limits on the RPV couplings and gravitino masses. 
We use the $CL_s$ method~\cite{Read:2002hq,Read_CLs} because of its robustness in the face of low backgrounds with possible downward fluctuations. 

In this calculation, the NFW DM halo profile and 'med' propagation parameters are assumed to be correct, and we have focused on the uncertainties in antideuteron production. 
Ideally, the uncertainties related to the propagation and the DM distribution should also be taken into account in the $CL_s$ calculation. 
These uncertainties are, however, difficult to estimate with our current knowledge, and in the literature they are typically handled by giving upper and lower limits corresponding to the 'max' and 'min' propagation models and extremal halo profile cases. 
The antideuteron fluxes from the 'min' and 'max' model typically differ by $\sim2$ orders of magnitude, with the 'med' model lying somewhere in between -- more or less in the middle for kinetic energies $T\sim10^{-1}$~GeV/n, and increasingly closer to the 'max' model for increasing energies; see {\it e.g.}~Ref.~\cite{Dal:2012my}.

\subsection{The $CL_s$ method}
Given a test-statistic $X$ that is constructed to increase monotonically for increasingly signal-like events, the $CL_s$ confidence is given by
\beq \label{eq:CLs}
CL_s = \frac{P_{s+b}(X \leq X_{\rm obs})}{P_b(X \leq X_{\rm obs})},
\eeq
where $X_{\rm obs}$ is the value of the test-statistic measured by an experiment, and $P_{i}(X \leq X_{\rm obs})$ is the probability of having $X \leq X_{\rm obs}$ given that hypothesis $i$ is true. 
The subscripts $b$ and $s+b$ correspond to the background and signal+background hypotheses, respectively.
A signal can be considered excluded at a confidence level $CL$ when
\begin{equation}
1-CL_s \geq CL .
\end{equation}

We calculate the $CL_s$ limits using the best fit point in $p_0$ and the hadronization parameters, given in Sec.~\ref{sec:fitresults}, henceforth referred to as the nuisance parameters.
For an assumed value of the RPV coupling $\lambda$ and gravitino mass $m_{\tilde G}$, we use the test-statistic
\begin{equation} \label{eq:Qstat}
Q(\lambda, m_{\tilde G}) = e^{-s_{\rm tot}} \prod_{i=1}^{N_{\rm chan}} \left(1+\frac{s_i}{b_i} \right)^{n_i},
\end{equation}
where the product  is taken over all bins and channels of an experiment, $s_i=s_i(\lambda, m_{\tilde G})$ and $b_i$ are the expected number of signal and background events in bin $i$, $s_{\rm tot} = \sum_i s_i$ is the expected total number of signal events,
and $n_i$ is the observed number of events in bin $i$.

Ideally, one should also include the nuisance parameter uncertainties in the calculation, as these tend to weaken the limits somewhat. Instead of using $Q(\lambda, m_{\tilde G})$ one would then use a test statistic such as the profile likelihood ratio,
\begin{equation}
q(\lambda, m_{\tilde G}) = \frac{{\mathcal L}(\lambda, m_{\tilde G},\hat{\hat{\theta}}_i)}{{\mathcal L}(\hat \lambda, \hat m_{\tilde G},\hat\theta_i)},
\end{equation}
formed from the ratio of the likelihood for the choice of $\lambda$ and $m_{\tilde G}$ and the corresponding best fit values of the nuisance parameters $\hat{\hat{\theta}}_i$, 
and the likelihood of the best fit point for $\lambda$, $m_{\tilde G}$ and the nuisance parameters $\theta_i$.
The relevant likelihood would then be
\begin{equation}
{\mathcal L}(\lambda, m_{\tilde G},\theta_i) = {\mathcal L}_{\rm cosmic}(\lambda, m_{\tilde G},\theta_i) \cdot {\mathcal L}_{\rm collider}(\theta_i),
\end{equation}
where ${\mathcal L}_{\rm collider}(\theta_i)$ is the likelihood function corresponding to the $\chi^2$ distribution found in our tuning of the nuisance parameters against collider experiments, 
while ${\mathcal L}_{\rm cosmic}(\lambda, m_{\tilde G},\theta_i)$ is the likelihood for a given observation at one of the cosmic ray experiments for the particular values of $\lambda$, $m_{\tilde G}$, and $\theta_i$.

Unfortunately,  the likelihood maximization is too computationally expensive to perform for every gravitino mass and coupling as it involves generating large samples of antideuterons for many different nuisance parameters, where each sample takes days on a single CPU due to the inefficiencies in the per-event coalescence model. From a test based on the $\bar{U}_1\bar{D}_1\bar{D}_2$ operator for a specific gravitino mass of 200~GeV, we estimate that including the nuisance parameter uncertainties can weaken the limits we set by approximately 20\% in the experiments studied here.

\subsubsection{Limits with single bin counting experiments}
For single bin counting experiments such as BESS and GAPS, the signal exclusion confidence in the $CL_s$ scheme can be simplified to~\cite{Read_CLs}
\beq
CL = 1-\frac{\sum_{n=0}^{n_{\rm obs}} \frac{e^{-(b+s)}(b+s)^n}{n!}}{\sum_{n=0}^{n_{\rm obs}} \frac{e^{-b}b^n}{n!}} ,
\eeq
where $s$ and $b$ are the expected number of signal and background events, and $n_{\rm obs}$ is the number of observed events.
For the case of zero observed events, which we assume here, this expression further simplifies to
\beq
CL = 1-e^{-s} ,
\eeq
which is independent of the background expectation. 

\subsubsection{Setting $CL_s$ limits for AMS-02}

For AMS-02, we have 2 bins, corresponding to the TOF and RICH detector sensitivity regions. 
The background expectations $b_1$ and $b_2$ for these detectors are known, and given a value for the RPV coupling in question, we can find the corresponding signal expectations, $s_1$ and $s_2$.
We assume $n_1=0$ observed events in the TOF detector and $n_2=1$ observed event in the RICH detector, and calculate $Q_{\rm obs}$ according to Eq.~\eqref{eq:Qstat}.

We then proceed by calculating $P_{s+b}(Q \leq Q_{\rm obs})$ and $P_{b}(Q \leq Q_{\rm obs})$.
This is done by finding all pairs of $(n_1, n_2)$ such that the corresponding value of $Q$ calculated from Eq.~\eqref{eq:Qstat} fulfills $Q \leq Q_{\rm obs}$. We denote this set of pairs $\mathcal C$.
Assuming first that the signal+background hypothesis is true, and that the number of events in bin $i$ follows a Poisson distribution with expectation value $s_i+b_i$, 
the probability of observing exactly $n_1$ and $n_2$ events at TOF and RICH, respectively, is given by a product of two Poisson distributions,
\beq \label{eq:probEvents}
P_{s+b}(n_1,n_2) = \frac{(s_1+b_1)^{n_1}}{n_1!} e^{-(s_1+b_1)} \frac{(s_2+b_2)^{n_2}}{n_2!} e^{-(s_2+b_2)}.
\eeq
The probability $P_{s+b}(Q \leq Q_{\rm obs})$ is then given by the sum of all probabilities $P_{s+b}(n_1,n_2)$ where $(n_1, n_2)$ satisfies $Q \leq Q_{\rm obs}$, {\it i.e.}
\beq
P_{s+b}(Q \leq Q_{\rm obs}) = \sum_{(n_1, n_2) \in {\mathcal C} } P_{s+b}(n_1,n_2).
\eeq

The procedure for calculating $P_{b}(Q \leq Q_{\rm obs})$ is identical, but using $s_1 = s_2 = 0$ in Eq.~\eqref{eq:probEvents}.
Inserting these results in Eq.~\eqref{eq:CLs}, we then find the $CL_s$ confidence value for the given value of the coupling.

The 95\% $CL_s$ confidence limit on the flux or coupling is finally found by varying the flux or coupling until the value for which $CL_s = 1 - 0.95$ is found.

\subsection{Current BESS limits}
\begin{figure}
\includegraphics[width=0.5\textwidth]{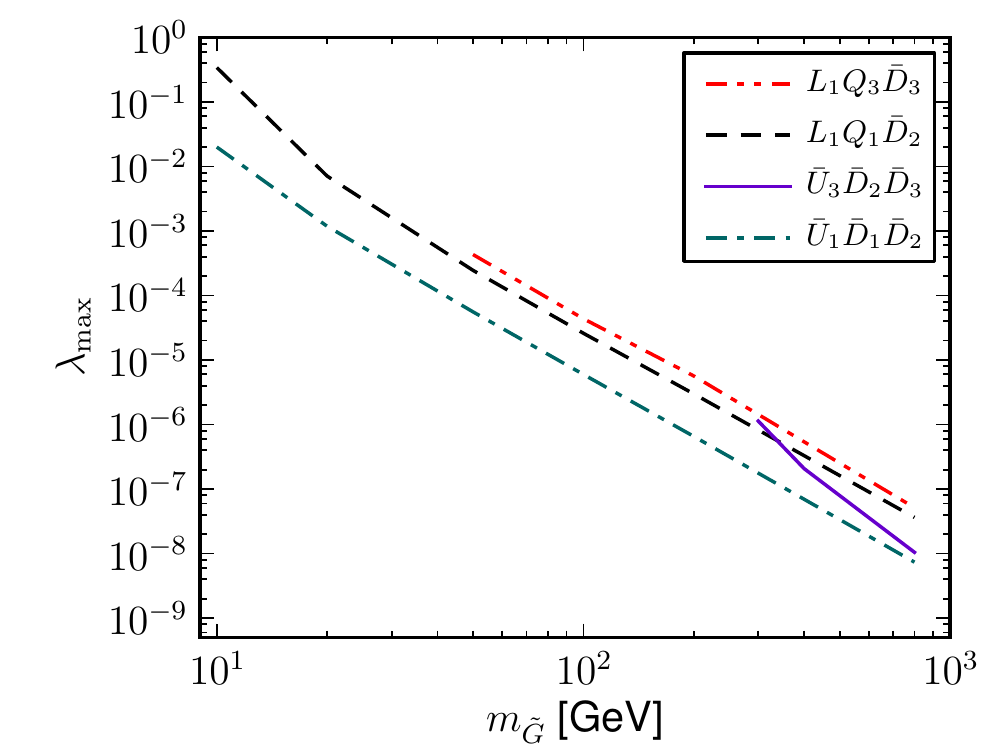}
\caption{Upper limits on various RPV couplings $\lambda$ as a function of gravitino mass from BESS antideuteron searches. The colours show different RPV couplings. All limits are for the 'med' propagation model and NFW DM halo profile.}
\label{fig:currentlimit}
\end{figure}

Figure~\ref{fig:currentlimit} shows the limits we can currently set on various RPV couplings from the non-observation of antideuterons by BESS. We see that, as expected, overall limits on the $\bar U\bar D\bar D$  couplings are stronger than for $L Q\bar D$, although some of this effect is, as pointed out above, from the shorter lifetime of the gravitino, and not directly connected to the (anti)quark content of the operator.

The limits on the $L_1 Q_1\bar{D}_2$ and $\bar{U}_1 \bar{D}_1\bar{D}_2$ operators from antideuterons at BESS are somewhat weaker than the existing limits found using antiproton data from PAMELA in Ref.~\cite{Bomark:2009zm}. 
For the $L_1 Q_3\bar{D}_3$ and $\bar{U}_3\bar{D}_2\bar{D}_3$ operators, the limits disappear when reaching the kinematical thresholds of the corresponding gravitino three-body decays.

The reader is warned to keep in mind that the antiproton limits in Ref.~\cite{Bomark:2009zm} were calculated using {\tt Pythia 6.4} for event generation and {\tt GALPROP} for cosmic ray propagation, with somewhat different propagation model parameters than used in the present study. Differences in the hadronization and propagation model can lead to differences in limits, and this should be kept in mind  when comparing the antiproton and antideuteron limits.

\subsection{Expected AMS-02 limit}
\begin{figure}
\includegraphics[width=0.5\textwidth]{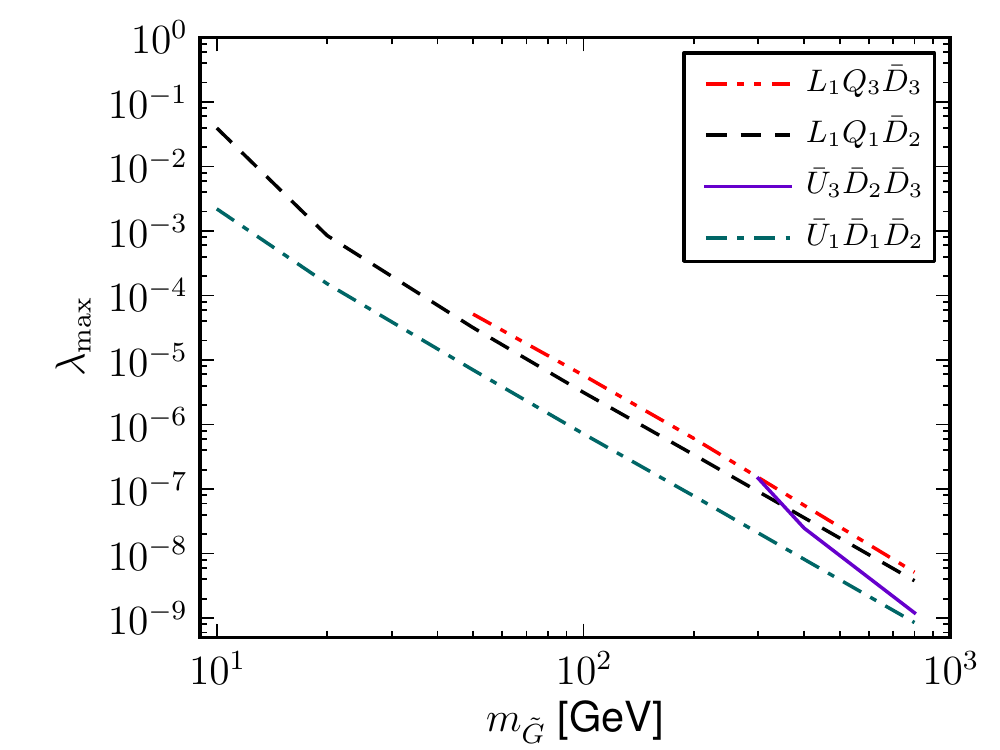}
\caption{Expected achievable upper limits on various RPV couplings $\lambda$ as a function of gravitino mass from the AMS-02 experiment. The limit assumes zero and one event observed in the TOF and RICH energy ranges, respectively. The colours show different RPV couplings. All limits are for the 'med' propagation model and NFW DM halo profile.}
\label{fig:expectedlimitAMS}
\end{figure}

In Fig.~\ref{fig:expectedlimitAMS} we show the expected achievable limit on the same RPV couplings from the AMS-02 experiment under the assumption of one observed event, consistent with the expected background. We see an overall significant strengthening of the bounds compared to the bounds from BESS.
The expected limit on the $L_1 Q_2\bar{D}_2$ operator is somewhat weaker than the current limit from antiprotons in Ref.~\cite{Bomark:2009zm} for low gravitino masses, but roughly equal for gravitino masses above a few hundred GeV.
For the $\bar{U}_1 \bar{D}_1\bar{D}_2$ operator, the expected limit from AMS-02 is a factor $\sim1.5$ stronger than the current limit for $m_{\tilde{G}}=10$~GeV, increasing to a factor $\sim4$ for gravitino masses in the 100~GeV range.

\subsection{Expected GAPS limit}
\begin{figure}[h!]
\includegraphics[width=0.5\textwidth]{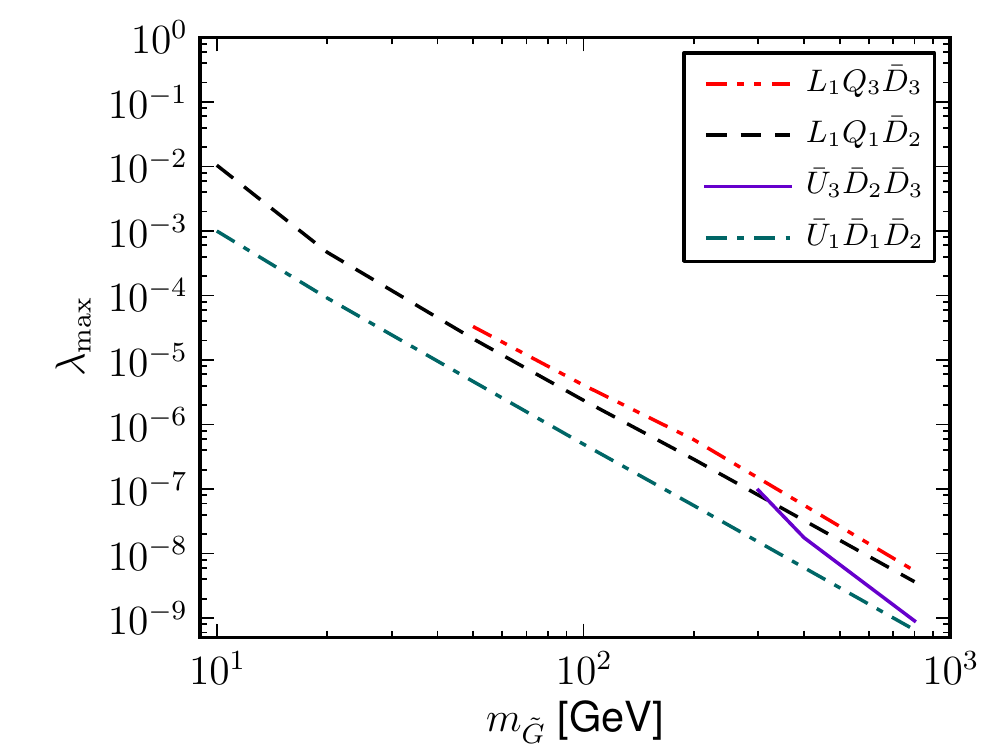}
\caption{Expected achievable upper limits on various  RPV couplings $\lambda$ as a function of gravitino mass from the LDB+ flight of the GAPS experiment. The limit assumes zero events observed. The colours show different RPV couplings. All limits are for the 'med' propagation model and NFW DM halo profile.}
\label{fig:expectedlimitGAPS}
\end{figure}

The prospective limits on RPV couplings from the GAPS experiment are shown in Fig.~\ref{fig:expectedlimitGAPS}.
We see that the expected limits on the RPV couplings from GAPS are a factor 2--4 stronger than the expected limits from AMS-02 for the lowest gravitino masses, but are roughly equal for gravitino masses approaching 800~GeV.
This is not unexpected, as the GAPS sensitivity range lies at lower energies than the AMS-02 ranges, and the peak in the antideuteron spectrum moves towards higher energies with increasing gravitino masses.

\section{Conclusions}
In this paper we have studied the influence of coalescence and hadronization parameters on the yield of antideuterons in dark matter decay/annihilation. By fitting to relevant collider data we provide a recommended set of parameter values for the {\tt Herwig++} event generator in Table~\ref{tab:fitresults}, with corresponding uncertainties that can be used for error propagation.

With the tuned formation model we have set bounds on trilinear RPV couplings in scenarios with gravitino dark matter using the results of antideuteron searches with the BESS experiment. 
These are of the same order of magnitude as limits from antiproton searches with PAMELA. 
We find large differences between different types of operators, the antideuteron searches are far more constraining---an order of magnitude---for the baryon number violating operators, and are more constraining for operators with light quarks. 
We have further investigated the potential of AMS-02 and GAPS to improve on these limits, and found that future results can strengthen current limits by a factor of $\sim 4$ for the baryon number violating operators in the whole range of gravitino masses considered (10--800 GeV). This will significantly improve on the current best limits for trilinear RPV couplings that come from antiproton searches.

\acknowledgments

LAD and ARR would like to thank Andreas Papaefstathiou for valuable assistance with the internal workings of {\tt Herwig++} and moral support, 
and Philip von Doetinchem for clarifying details in the calculation of expected events in the experiments.
We would also like to express our gratitude to Alex Read for his patience in explaining statistics to theorists. 
The CPU intensive parts of this work were performed on the Abel Cluster, owned by the University of Oslo and the Norwegian metacenter for High Performance Computing (NOTUR), and operated by the Research Computing Services group at USIT, the University of Oslo IT-department. The computing time was given by NOTUR allocation NN9284K financed through the Research Council of Norway.


\bibliographystyle{plain}	

\end{document}